\documentclass[preprint,aps,nofootinbib,showkeys,showpacs,12pt]{revtex4-1}

\usepackage{graphicx}
\usepackage{epsfig}
\usepackage{hhline}
\usepackage{amsmath}  
\usepackage{color}
\usepackage{amssymb}

\begin{document}

\title{What Does Low Energy Physics Tell Us About The Zero Momentum Gluon Propagator}

\author{P. Costa$^1$}
\author{O. Oliveira$^1$ }
\author{P. J. Silva$^1$ }
\affiliation{$^1$ Departamento de F\'{\i}sica, Universidade de Coimbra, P-3004-516 Coimbra, Portugal}

\begin{abstract}
The connection between QCD, a nonlocal Nambu--Jona-Lasinio type model and the Landau gauge
gluon propagator is explored. This two point function is parameterized by a functional form which is compatible with Dyson-Schwinger 
and lattice QCD results. 
Demanding the nonlocal model to reproduce the experimental values for the pion mass, the pion decay constant,
$\Gamma_{\pi \rightarrow \gamma\gamma}$ and the light quark condensate we conclude that low energy physics does not distinguish
between the so-called decoupling and scaling solutions of the Dyson-Schwinger equations. This result means that, provided that the model
parameters are chosen appropriately, one is free to chose any of the above scenarios. 
Furthermore, the nonlocal Nambu--Jona-Lasinio quark model considered here is chiral invariant and satisfies the GMOR relation
at the 1\% level of precision.
\end{abstract}

\maketitle


\newpage
The usual approach to Quantum Chromodynamics (QCD) defines the Green's 
function generation function  \textit{\`a la} Faddeev-Popov (FP) \cite{FaPo67}.
This approach allows the investigation of the perturbative solution of QCD. 
However, in what concerns nonperturbative phenomena, like confinement and 
chiral symmetry breaking, the FP trick is ill defined as it assumes a unique 
solution of the gauge fixing condition on  each gauge orbit. 
Gribov demonstrated \cite{Gribov}, for the Coulomb and Landau gauges, 
that in each set of gauge related configurations there are more than one 
configuration, the Gribov copies, satisfying the gauge fixing condition. The 
study of Gribov was confirmed by many lattice simulations \cite{silva04}, 
which also investigate how the copies modify the Green's functions. 
Furthermore, there is a no-go theorem \cite{singer78} stating that the 
problem of the copies is common to all local 
gauge fixing conditions.  

Gribov tried to solve the problem of the copies by changing the 
functional integration space, replacing 
the  hyperplane of transverse configurations,
$\Gamma ~ = ~ \left\{ A: ~ \partial \cdot A = 0 \right\}$,
by the first Gribov region $\Omega$:
the subset of $\Gamma$ with a positive defined Faddeev-Popov operator. 
This change of integration space modifies
the tree level gluon propagator in the infrared region and, 
according to Gribov, at zero momentum the propagator vanishes.
However, the problem of building a nonperturbative generating functional 
for QCD was not solved, as 
$\Omega$ is not free of Gribov copies.

In \cite{Zwanziger89,Zwanziger90} the restriction of the integration 
space to $\Omega$ was implemented via a modification of
the QCD action. The Gribov-Zwanziger (GZ) action includes extra bosonic 
and fermionic fields, 
breaks softly BRST \cite{Zwanziger90,Dudal08} and is perturbatively 
renormalizable. 
Moreover, the GZ action also predicts a vanishing zero momentum gluon 
propagator \cite{Zwanziger91}. This result suggested a gluonic confinement 
criteria as discussed in \cite{Zwanziger89,Zwanziger94}. 
Further extensions of the GZ action, including possible condensates, were 
investigated  in \cite{Dudal08} leading to the so-called refined 
Gribov-Zwanziger (RGZ) action. If the GZ action predicts a $D(0) = 0$, 
according to the RGZ action $D(0) \ne 0$ \cite{DudalOliveira10}.

In the recent years there has been a considerable effort to compute the 
gluon and ghost propagators from first principles, namely using 
Dyson-Schwin\-ger equations (DSE) and the lattice formulation of QCD. 
Considerable attention was given to the infrared propagators and to 
the value of $D(0)$. 

DSE computations start from the Faddeev-Popov quantization procedure. 
The solutions follow into two categories: the scaling solution 
\cite{LeSm02,ReviewScaling} which has $D(0) = 0$ and is compatible 
with the Gribov-Zwanziger confinement criterium; the decoupling 
solution \cite{AgNa03,BiPapa09,orsay} which predicts a $D(0)\neq0$, 
related to a dynamical generated  gluon mass $M(p^2)$. 

Wilson action 4D lattice QCD simulations for the gluon and ghost 
propagators have been performed for SU(2) and SU(3) gauge groups
\cite{Leinweber,Braz,Trento,qnp,Cucc07, CucPRL,Bo09,OlSil09}.
The gluon propagator has been computed for volumes as large as (27 fm)$^4$ 
for SU(2) \cite{CucPRL} ($a \sim 0.21$ fm), and (16 fm)$^4$ for 
SU(3) \cite{Bo09}  ($a \sim 0.17$ fm). 
The raw data shows a decreasing $D(0)$ with the lattice volume but no 
infrared suppression of $D(p^2)$ has been observed. Furthermore, naive 
extrapolations of $D(0)$ to the infinite volume \cite{Cucc07} do not 
provide a clear answer about the infinite volume value for $D(0)$.

In \cite{CucPRL}, inequalities relating the zero momentum gluon propagator 
with an average absolute value color-magnetization, which allow for a scaling
analysis with the volume, were derived. SU(2)  simulations \cite{CucPRL} 
give a finite and nonvanishing $D(0)$ 
in the limit $V \rightarrow + \infty$. A similar analysis for SU(3) 
\cite{OlSil09}, although using smaller volumes, give $D(0) = 0$ in 
the same limit. The reasons for this difference are still not understood.

In \cite{OlSi09b} the authors considered  ratios of lattice propagators, 
which, hopefully, suppress the finite volume effects. The analysis of 
the ratios point towards a vanishing $D(0)$ in the infinite volume limit.

In conclusion, the present situation gives ambiguous results for the value 
of $D(0)$. This question is of great theoretical relevance as it validates, 
or not, possible gluon confinement mechanisms. 
Any other way of estimating $D(0)$ independently is welcome. In this work 
we link $D(0)$ with low energy physics, i.e. the physics associated with pions. 
In order to connect the infrared gluon propagator with low energy 
phenomenology, an effective low energy chiral quark model of the 
Nambu-Jona-Lasinio type is built.


In the QCD Lagrangian, the interaction between quarks and gluons reads
\begin{equation}
  \mathcal{L}_{\overline\psi\psi A} ~ = ~ g \, \overline\psi \, \gamma^\mu \, A^a_\mu \, \frac{\lambda^a}{2} \, \psi
  \, .
\end{equation}  
In the generating functional, expanding
the exponential term containing
$ \mathcal{L}_{\overline\psi\psi A}$ to second order in $g$ and integrating the gluon fields,  the theory becomes 
an effective nonlocal fermionic theory. For one flavor QCD, after color and spin Fierz transformations, 
assuming that vector and axial vector currents play a secondary role in the dynamics, and taking into account only 
the color singlet bilocal currents $J (x,y) =  \overline \psi^i (x) \psi^i (y)$, $J_5 (x,y)  =  \overline \psi^i (x) \gamma_5 \psi^i (y)$
($i$ stands for color indices) one gets the effective action
\begin{eqnarray}
 S[\overline\psi,\psi]  &=&  \!\!\!\int d^4x d^4y \Big\{  \overline\psi (y) \, \delta (y-x) \, \left( 
           i \gamma^\mu \partial_\mu - m \right) \psi (x) \nonumber \\
  &+& \!\!\frac{g^2}{8}  \, J (x,y) D (x -y)   J (y,x)         -  
                \frac{g^2}{8} \, J_5 (x,y)  D (x -y)  J_5 (y,x) \Big\}   .
   \label{EffAction}
\end{eqnarray}
This action is invariant under global chiral symmetry, i.e. 
$   \psi \longrightarrow e^{i \theta \gamma_5} \psi$ and  
$   \overline\psi \longrightarrow  \overline\psi e^{-i \theta \gamma_5} $ ,
provided that $m = 0$, and is sensible to the details of the gluon propagator $D(x-y)$. 
Extending (\ref{EffAction}) to  include various quark flavors 
is straightforward. In the following it will be assumed SU(2)-flavor symmetry and the
current quark masses will be written as $m_u = m_d = m_q$.

The action (\ref{EffAction}) is a nonlocal version of the Nambu--Jona-Lasinio (NJL) action \cite{NJL}. 
The NJL model gives a good description of low energy hadron phenomenology and it is expected that the nonlocal version should do a
 similar job, with the extra bonus of being sensible to the gluon propagator. 
The NJL model is nonrenormalizable and needs to be regularized by the introduction of a finite cut-off $\Lambda$. 
For the local version, where $D(x-y) \propto \delta (x-y)$, typical cut-off values range from 600 MeV up to 800 MeV. 
For the model described by Eq.  (\ref{EffAction}), besides the regularization, one also needs to know the gluon form factor
$D(x-y)$. 

\begin{figure}[t]
  \centering
		\epsfig{file=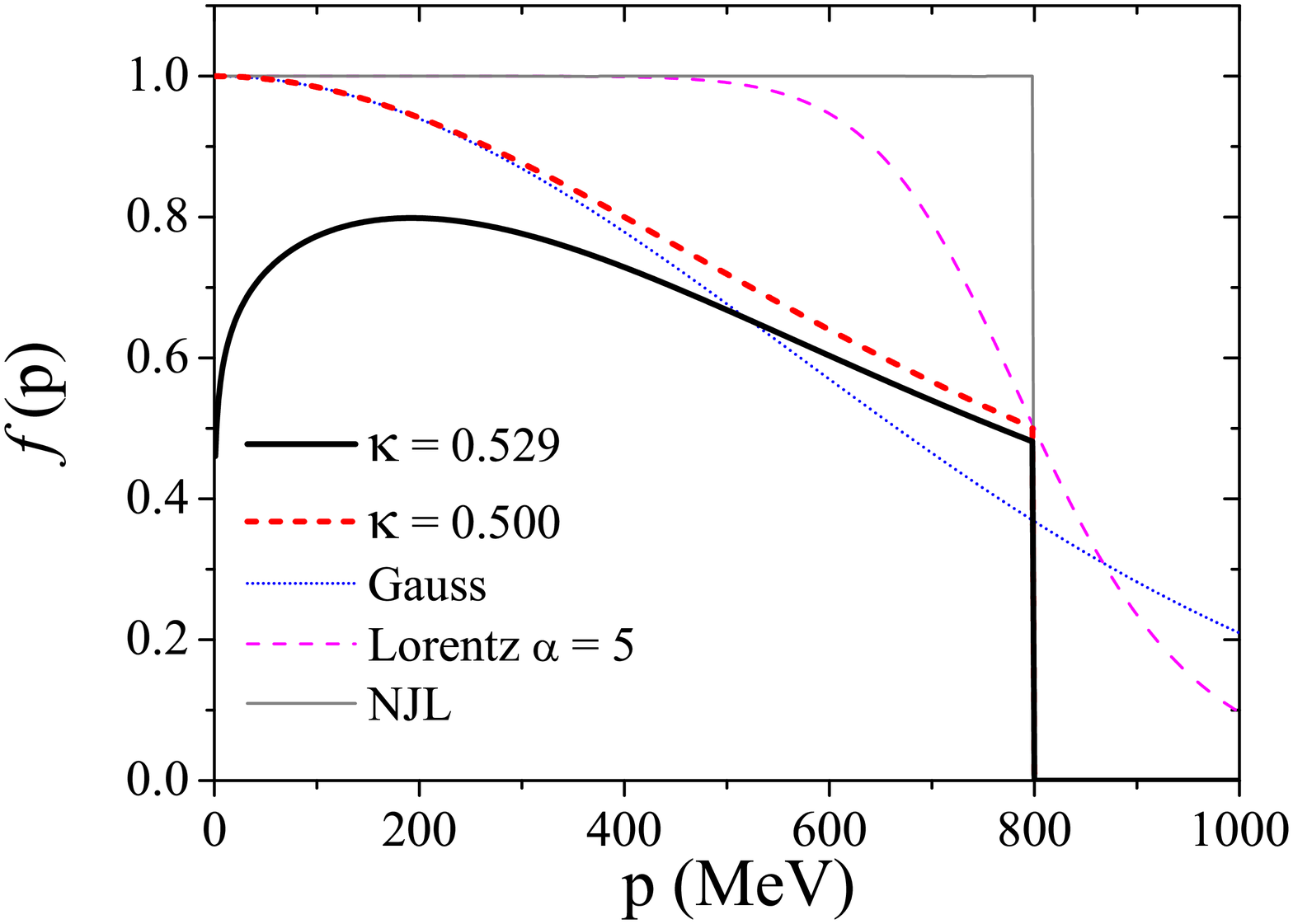,width=10.0cm,height=6.5cm}
  \caption{Form factors as a function of $p$.  The figure includes typical form factors used in  previous studies --
                  see, for example,  \cite{Grigorian:PPNL_2007} and references there in. \label{fig:f_forma}}
\end{figure}

First principles calculations of the gluon propagator have been performed using lattice QCD and DSE. 
The momentum space propagator 
\begin{equation}
        D(p^2)\,  = \, Z \, \frac{ \left(  p^2 \right)^{2 \kappa - 1}}{\left( p^2 + \Lambda^2_{QCD} \right)^{2 \kappa}} 
      \label{GlueProp}
\end{equation}
is able to describe both the scaling ($\kappa>0.5$) and decoupling ($\kappa=0.5$) infrared DSE solutions and the lattice data up to 
$p\sim800 MeV$ -- see Refs. \cite{Braz,Trento} for discussions.
In Eq. (\ref{GlueProp}), $\Lambda_{QCD}$ stands for an infrared mass scale. 

Let us define the dimensionless form factor in momentum space as
\begin{equation}
f({p^2}) ~ = ~ \Lambda^2 D(p^2) ~ = ~ \frac{\Lambda^2}{{p}^2} \, 
         \left(\frac{{p}^2}{{p}^2 + \Lambda^2_{QCD}}\right)^{2\kappa} \theta(\Lambda-{p}) \, .
\label{form_factorS}
\end{equation}
The constant $Z$  in Eq. (\ref{GlueProp}) will be included in the 
definition of the coupling constant $G$, which multiplies the quark 
currents of the nonlocal theory.  $G$ carries the dimension of a length
squared.  In $f(p^2)$, $\Lambda$ is the cut-off. In a first step
we assume $\Lambda_{QCD} = \Lambda$. 
The form factor $f(p^2)$, together with typical form factors 
considered in the literature,  is shown in Fig. \ref{fig:f_forma}.

The model has four parameters $\{G, m_q, \kappa, \Lambda \}$ to be
determined in vacuum. Given that Eq. (\ref{GlueProp}) reproduces 
the lattice data for $p < 800$ MeV, we take $\Lambda = 800$ MeV.
For each $\kappa$, $G$ and $m_q$ are adjusted to reproduce the 
experimental pion mass $M_\pi = 139.57$ MeV
and decay constant $f_\pi = 92.4$ MeV.

We also take in consideration the light quark condensate 
$\langle\bar {q}q\rangle$. At a renormalization scale of  $\mu=1$ GeV, 
$\langle\bar {q}q\rangle^{1/3}=-240$ MeV, while at $\mu=2$ GeV it
becomes $\langle\bar{q}q\rangle^{1/3}=-270$ MeV 
\cite{Jamin:2002ev,Bordes:2010wy}. The difference is mainly due to
the running of $\hat{m}$ from 5.5 MeV, for $\mu=1$ GeV, to 4.1 MeV 
if $\mu=2$ GeV - see Ref. \citep{pdb}.  
Furthermore, due to phenomenological reasons, 
we will restrict our study to constituent quark masses $M_q$ in the range 
$300$ -- $400$ MeV, which are typical values of $M_q$ in this type of models.
The decay width for $\pi \longrightarrow \gamma\gamma$ will be used
to distinguish the different sets of parameters.

Our nonlocal model belongs to a class of models which have been 
explored by previous authors
\cite{Schmidt:PRC_50_1994,Blaschke:NPA_592_1995,Blaschke:IJMPA_16_2001} 
and we will not discuss how to compute the different quantities. 
The technical details can be found
in \cite{Schmidt:PRC_50_1994}. 


We start the discussion of the results
with a technical comment on the NJL--type model calculations.  
The fermionic action of NJL--type models has ultraviolet divergences 
and requires a regularizing cut-off \cite{costa:PLB_577_2003}. 
However, not all the integrals are divergent and some authors
chose to regularize only the action and the divergent integrals: 
for the finite integrals they remove
the cut-off. This has an important impact
on $\Gamma_{ \pi \rightarrow \gamma\gamma}$. Indeed, for the local NJL model
the width computed with all integrals regularized give, typically, only
about 70\% of the experimental numbers (see, for example, 
\cite{Blaschke:NPA_592_1995,costa:PLB_577_2003_2}).
 In our calculation the cut-off $\Lambda = 800$ MeV
was used at all the stages of the computation.

We would like to remember the reader that our work has two main goals. 
On one side we would like to define a nonlocal model for low energy 
physics which is, as much as possible, rooted in first principle 
computations. This motivates our choice for the form factor. 
On the other side, we would like to connect low energy physics with 
the behavior of the gluon propagator $D(p^2)$ in the infrared region 
and, in particular, its value for $p=0$. 

\begingroup
\begin{table}
	\begin{center}
  	\begin{tabular}{|c|c|c|c|c|c|c|}
    	\hhline{|======|}
      	& $m_q$ [MeV] & $M_q$ [MeV]& $-\left\langle \bar{q}q \right\rangle^{1/3}$ [MeV]& $\,\,G\Lambda^2\,\,$ 
      	&	$\Gamma_{\pi\gamma\gamma}$  [eV]  \\
      \hhline{|======|}
      $\kappa = 0.40  $ & 4.093       & 295.8       & 273.3		& 3.781		& 2.06  \\
      \hline
      $\kappa = 0.45  $ & 4.146       & 325.6       & 272.2		& 4.946		& 3.22  \\
      \hline
      $\kappa = 0.50  $ & 4.187       & 360.5       & 271.4		& 6.441		& 5.44  \\
      \hline
      $\kappa = 0.529 $ & 4.205       & 383.6       & 271.1		& 7.491		& 7.79  \\
     \hline
      $\kappa = 0.548 $ & 4.214       & 400.1       & 271.0   & 8.263		& 10.19  \\
     \hline
      $\kappa = 0.60  $ & 4.231       & 453.3       & 271.4   & 10.776	& 26.20  \\
     \hline
		\end{tabular}
		\caption{A collection of parameters which reproduce the 
experimental $M_\pi$ and $f_\pi$ for a cut-off $\Lambda=\Lambda_{QCD}=800$ 
MeV. The results differ essentially on the value of the decay width 
$\Gamma_{ \pi \rightarrow \gamma\gamma}$ (last column). Recall that the 
experimental width is 7.78(56) eV. \label{TableI}}
	\end{center}	
\end{table}
\endgroup

In what concerns $M_\pi$ and $f_\pi$, for a cut-off of $\Lambda = 800$ MeV, 
it turns out that the experimental values can be reproduced by a large 
set of parameters, including $\kappa$ above and below 0.5 -- see 
Table \ref{TableI}. The values on Table \ref{TableI} show that the 
quark condensate is almost independent of $\kappa$ and always close 
to the expected value $\langle\bar{q}q\rangle^{1/3}=-270$ MeV.
It is important to point out that this model is 
consistent with the Gell-Mann--Oakes--Renner relation (GMOR) 
\begin{equation}\label{gmor}
 	M_\pi^2 f_\pi^2=-2m_q\langle\bar{q}q\rangle_0\,,
\end{equation}
preserving chiral low-energy theorems and current algebra 
relations \cite{Grigorian:PPNL_2007}. 
As matter of fact, we can calculate the GMOR value for the current quark mass
\begin{equation}\label{gmor_mq}
 	m_q^{GMOR}=-\frac{M_\pi^2 f_\pi^2}{2\langle\bar{q}q\rangle_0}\,,
\end{equation}
as an indicator of the validity of this low energy theorem. Taking for
example $\kappa=0.5$, $m_q^{GMOR}=4.159$ MeV which differs less then $1\%$ 
of the calculated value $m_q=4.187$ MeV.

\begin{figure}[t]
  \centering
		\epsfig{file=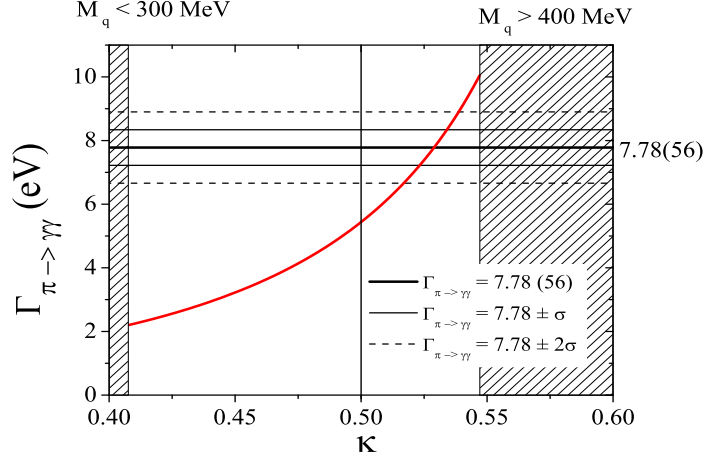,width=10.0cm,height=6.5cm}
  \caption{$\Gamma_{\pi\rightarrow\gamma\gamma}$ as a function of $\kappa$ for
                  $\Lambda = \Lambda_{QCD} = 800$ MeV.
The excluded regions, $\kappa< 0.41$ and $\kappa > 0.55$, are associated with a constituent quark mass outside
typical phenomenological values $M_q <  300$ MeV or $M_q > 400$ MeV.
See also Table \ref{TableI}.
                   \label{fig:dec1}}
\end{figure}

\begin{figure}[t]
  \centering
    \includegraphics[width=1.\textwidth]{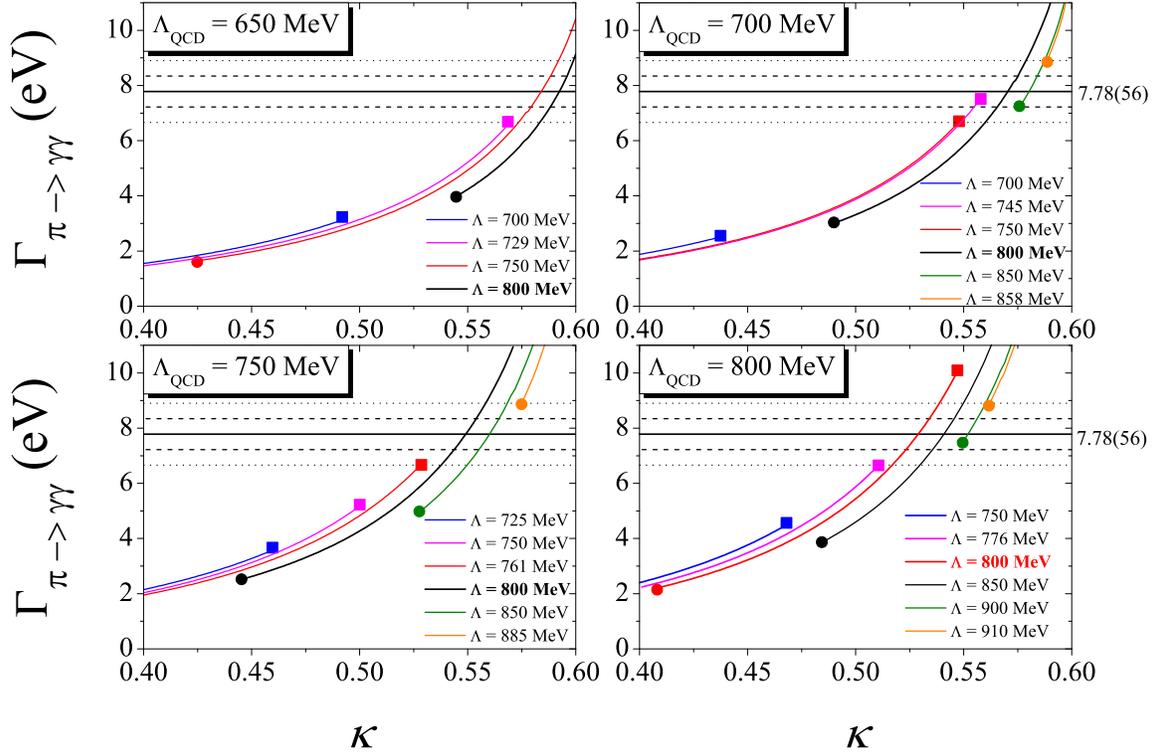}
  \caption{$\Gamma_{\pi \rightarrow \gamma\gamma}$ as a function of $\kappa$ for several values of $\Lambda$ and  $\Lambda_{QCD}$. The circles and squares mean $M_q=300$ MeV and  $M_q=400$ MeV respectively.    \label{fig:varios}}
\end{figure}

The width $\Gamma_{\pi \rightarrow \gamma\gamma}$, 
with an experimental value of $7.78\, (56)$ eV \cite{cleo}, 
distinguishes between the various scenarios for $D(0)$ -- see
Fig. \ref{fig:dec1}.
Clearly, the two pion decay points towards a $\kappa$ above 0.5
and a $D(0) = 0$. This conclusion is unchanged when $\Lambda \ne \Lambda_{QCD}$ 
and for other cut-off values  -- see Fig. \ref{fig:varios}. 

The results reported in Fig. \ref{fig:varios} show that to reproduce 
the experimental $\Gamma_{\pi \rightarrow \gamma\gamma}$, the cut-off 
$\Lambda$ and the infrared mass scale $\Lambda_{QCD}$ are correlated. 
The cut-off increases with $\Lambda_{QCD}$, with
$\Lambda$ and $\Lambda_{QCD}$ being of the same order of magnitude.
Moreover, for each $\Lambda_{QCD}$, there is only a narrow band of 
cut-off values which are able to reproduce the
experimental width. For example, for $\Lambda_{QCD} = 750$ MeV, $\Lambda$ 
has to be in the interval 761--885 MeV to reproduce 
$\Gamma_{\pi \rightarrow \gamma\gamma}$ up to two standard deviations.
Furthermore, $\kappa$ seems to decrease slightly when $\Lambda_{QCD}$ 
increases. Given the correlation between $\Lambda$ and $\Lambda_{QCD}$, 
this suggests that for higher values of the cut-off, in principle, 
one could reproduce the full set of 
observables for $\kappa = 0.5$ or smaller. However, since our 
nonlocal model does not take into account
the vector and axial-vector bilocal quark currents, that the 
light $J = 1$ spectrum starts at $\sim 800$ MeV,
an upper limit of $\Lambda$ around 800 MeV seems natural.
 If one wants to consider higher cut-off values,
one should also include the vector and axial-vector currents in the model.

\begin{figure}[t]
  \centering
		\epsfig{file=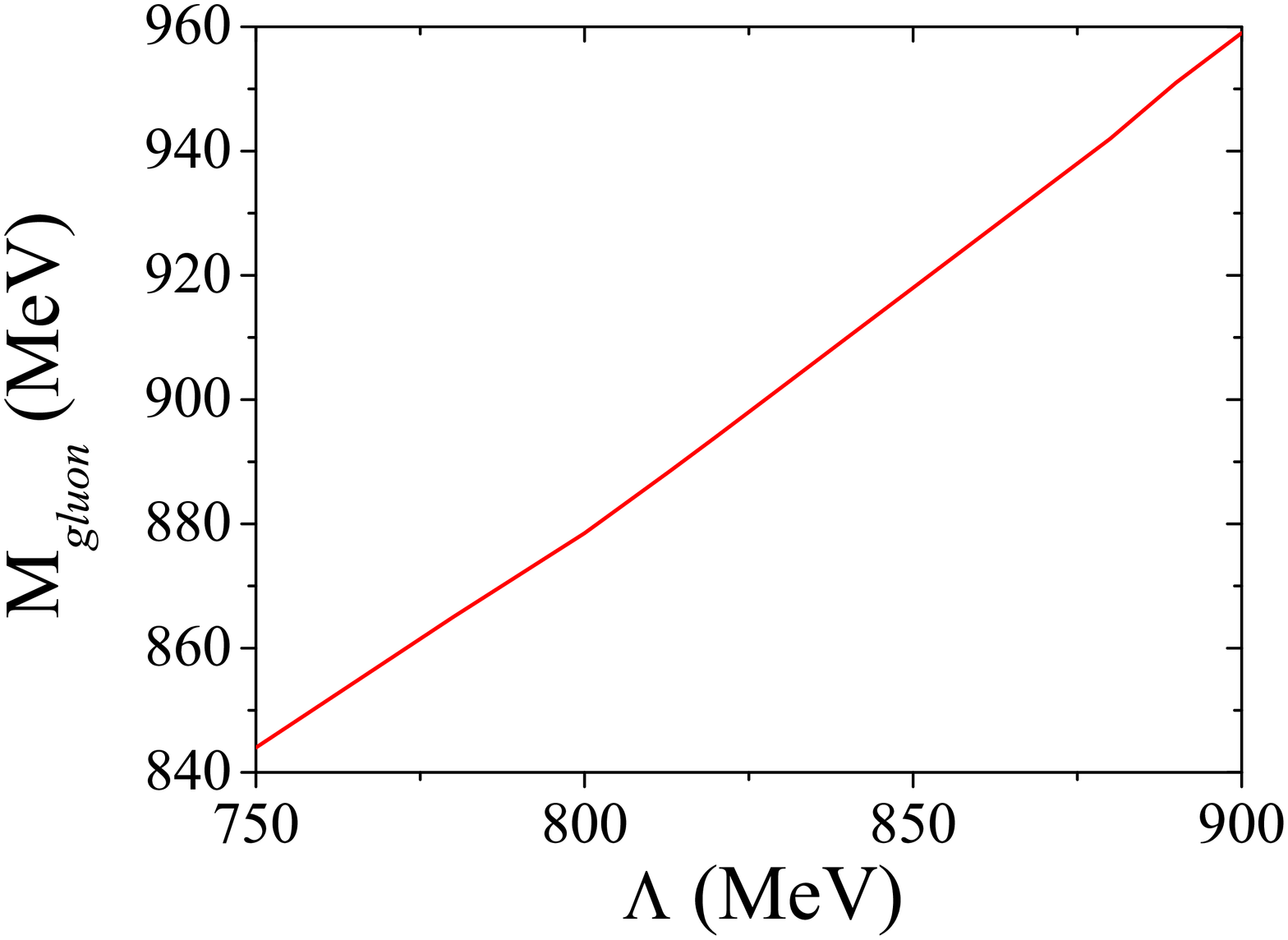,width=10.0cm,height=6.5cm}
  \caption{$M_{gluon} (\Lambda)$ required to reproduce the experimental
                $\Gamma_{\pi\rightarrow\gamma\gamma}$ .
                   \label{fig:dec2}}
\end{figure}

The decoupling type of propagator can be investigated setting
$\kappa = 0.5$ in Eq. (\ref{GlueProp}). 
Then the gluon propagator becomes
\begin{equation}
        D(p^2)\,  = \frac{ Z }{ p^2 + M_{gluon}^2} ;     \label{massive_fit}
\end{equation}   
$M_{gluon}$ takes the role of $\Lambda_{QCD}$ 
and can be interpreted as an effective gluon mass.
Following the same procedure as before, it follows that the model for 
$\kappa = 0.5$ also reproduces $M_\pi$, $f_\pi$ and 
$\Gamma_{\pi\rightarrow \gamma \gamma}$, provided the gluon mass is 
adjusted as a function of the cut-off $\Lambda$. Our results are
summarized in Table \ref{TableII} and Fig. \ref{fig:dec2}, for various 
cut-off values.

The effective gluon mass $M_{gluon} = \Lambda_{QCD}$ is within typical 
values found in the literature, see \cite{OlBic} and references 
there in,  but shows a strong dependence with the cut-off. 
$M_{gluon}$ is a linear function of $\Lambda$
-- see Fig. \ref{fig:dec2}. In what concerns the quark condensate, 
the model shows that $\langle \overline q q \rangle$
increases with $M_{gluon}$, i.e with the cut-off $\Lambda$. For 
$\kappa = 0.5$, to reproduce the experimental value of the 
condensate, i.e. to have $- \left\langle \bar{q}q \right\rangle^{1/3} = 270$ 
MeV, it turns out that the gluon mass is
$M_{gluon} = 878$ MeV for $\Lambda = 800$ MeV.

Finally, we would like to refer that the GMOR relation is satisfied always, i.e. for $\kappa > 0.5$ and $\kappa = 0.5$, within
the same level of precision.



In this paper, the connection between QCD and the NJL model is
explored to investigate possible links between the quenched gluon propagator
and low energy hadronic phenomenology.  The relation between the 
nonlocal model and QCD used is valid at the leading order level. 
It has been shown that nonlocal covariant extensions of the NJL model 
have the advantage, over the local theory, of the next-to-leading 
order corrections are relatively small \cite{Ripka}.
Therefore, we expect that, at least qualitatively, our results will 
still be valid beyond the leading order. 

Our main goal is to connect $D(p^2)$ with pion physics. The results 
discussed previously suggest that low energy physics is essentially 
blind to the gluon propagator at very low momenta. Indeed, the results 
show that one is free to use either the decoupling of scaling scenarios 
for $D(p^2)$, provided that the model parameters are chosen appropriately. 

For a scaling-type scenario, $M_\pi$ and $f_\pi$ are not sensitive to 
the deep infrared gluon propagator. Indeed both values can be reproduced 
for $\kappa$ above and below 0.5. 
However, $\Gamma_{\pi\rightarrow \gamma \gamma}$  is sensitive to $\kappa$. 
Curiously, for $\Lambda = \Lambda_{QCD} = 800$ MeV, the value of $\kappa$ 
which reproduces the experimental decay width, $\kappa = 0.529$, 
is in perfect agreement with the infrared exponent measured from 
lattice QCD in \cite{qnp,OlSi09b} 
and the estimate from time-independent stochastic quantization \cite{Zw02}. 
Of course, given the level of approximations used here, 
the good agreement is, probably, an happy coincidence.

\begingroup
\begin{table}[t]
	\begin{center}
  	\begin{tabular}{|c|c|c|c|c|c|c|}
    	\hhline{|======|}
      $\Lambda$ [MeV]	& $M_{gluon}$ [MeV] & $m_q$ [MeV] & $M_q$ [MeV] & $-\left\langle \bar{q}q \right\rangle^{1/3}$ [MeV] 
      	&	$\,\,G\Lambda^2\,\,$  \\
      \hhline{|======|}
      750 & 843.8       & 4.5       & 460.9		& 264.2		& 6.03  \\
      \hline
      800 & 878.7       & 4.2       & 409.8		& 271.6		& 5.39  \\
      \hline
      813 & 888.5       & 4.1       & 400.0	  & 273.5		& 5.27  \\
      \hline
      850 & 917.7       & 3.9       & 373.3		& 278.9		& 4.96  \\        
     \hline
      900 & 959.1       & 3.6       & 345.7		& 286.3		& 4.66  \\
     \hline
		\end{tabular}
		\caption{Summary of the results for $\kappa = 0.5$.}
		\label{TableII}
	\end{center}	
\end{table}
\endgroup

For a decoupling-type scenario, the "measured" effective gluon mass is within typical values.
From the values of the quark condensate, the massive gluon model seems to prefer a $M_{gluon} \sim 878$ MeV.

The parameterizations discussed in this work gather the essential ingredients of the scaling and decoupling solutions 
of the DSE and the lattice propagators for $p \in$ [50, 800] MeV. 
Besides the link with the infrared gluon propagator, the models analyzed also provide an effective nonlocal chiral quarks 
model and, in this sense, are a good starting point towards the construction of a more realistic nonlocal low-energy effective 
theory for QCD.

\vspace{-0.2cm}
\section*{Acknowledgments}
\vspace{-0.1cm}
The authors acknowledge financial support from F.C.T. under project CERN/FP/83644/2008 
and  (P.J.S.) grant SFRH/BPD/40998/2007. O.O acknowledges financial support from FAPESP.



\end{document}